\documentclass[12pt]{article}
\usepackage{amssymb,amsmath}
\usepackage{amsthm}
\usepackage{xcolor}
\usepackage{hyperref}
\theoremstyle{definition}

\def\d{\partial}
\def\f{\frac}

\newcommand*{\pd}
[2]{\mathchoice{\frac{\partial#1}{\partial#2}}
  {\partial#1/\partial#2}{\partial#1/\partial#2}
  {\partial#1/\partial#2}}

\newcommand*{\fd}
[2]{\mathchoice{\frac{\delta#1}{\delta#2}}
  {\delta #1/\delta#2}{\delta#1/\delta#2}{\delta#1/\delta#2}}
\newcommand{\ddx}[1]{\partial_x^{#1}}

\allowdisplaybreaks[3]
\begin{document}

\title{Weakly nonlocal Poisson brackets,\\
  Schouten brackets and supermanifolds}
\author{P. Lorenzoni$^1$, R. Vitolo$^2$}
\date{
  \framebox{
\begin{minipage}{9cm}
  \begin{center}
    \small\itshape Dedicated to J.S. Krasil'shchik \\
      on the occasion of his 70th birthday\\
      \normalfont Published in the \\
      Special Issue of Journal of Geometry and Physics
      \\
      \url{https://gdeq.org/SIJSK70}
\end{center}
\end{minipage}
}
}

\maketitle
\vspace{-7mm}
\begin{center}
  $^1$Dipartimento di Matematica\\
  Universit\`a di Milano-Bicocca
  \\
  email: \texttt{paolo.lorenzoni@unimib.it}
  \\[2mm]
  $^2$Dipartimento di Matematica e Fisica ``E. De Giorgi'',
  \\
  Universit\`a del Salento\\
  and Sezione INFN di Lecce\\
  via per Arnesano, 73100 Lecce, Italy
  \\
  e-mails:
  \texttt{raffaele.vitolo@unisalento.it}
\end{center}
\begin{abstract}
  Poisson brackets between conserved quantities are a fundamental tool in the
  theory of integrable systems. The subclass of weakly nonlocal Poisson
  brackets occurs in many significant integrable systems.  Proving that a
  weakly nonlocal differential operator defines a Poisson bracket can be
  challenging. We propose a computational approach to this problem through the
  identification of such operators with superfunctions on supermanifolds.
\end{abstract}

\section*{Introduction}

In the geometric approach to the integrability of PDEs a central role is played
by Poisson brackets \cite{ablowitz91:_solit,Dorfman:DSInNEvEq,
  Zakharov:WIsIn}. A system of PDEs
\begin{equation}
u^i_t = f^i(t,x,u^j,u^j_x,u^j_{xx},\ldots)\label{eq:5}
\end{equation}
with $n$ unknown functions $u^1$,\dots, $u^n$ (only two independent variables
$t$, $x$ for simplicity) admits a Hamiltonian formulation if there exists a
differential operator $P$ and a density
$\mathcal{H}=\int h \, dx$ such that
\begin{equation}
  u^i_t = P^{ij}\left(\fd{\mathcal{H}}{u^j}\right)
\end{equation}
where $P=(P^{ij})$ is a Hamiltonian operator, i.e. a matrix of differential
operators $P^{ij}= P^{ij\sigma}\partial_\sigma$, where
$\partial_\sigma = \partial_x\circ\cdots\circ \partial_x$ (total
$x$-derivatives $\sigma$ times), such that
\begin{equation}
  \{F,G\}_P    = \int \fd{F}{u^i}P^{ij\sigma}\partial_\sigma\fd{G}{u^j}\,dx
\end{equation}
is a \emph{Poisson bracket} (skew-symmetric and Jacobi). Such properties are
equivalent to the conditions
\begin{enumerate}
\item $P^*=-P$, skew-adjointness of $P$, and
\item $[P,P]=0$, where the (square) bracket is the \emph{(variational) Schouten
    bracket} of differential operators.
\end{enumerate}

Soon after the introduction of Hamiltonian operators in the theory of PDEs it
was observed that Hamiltonian operators for many PDEs were indeed nonlocal (or
pseudodifferential) operators. This fact was first described in \cite{Sok84}
for the Krichever--Novikov equation. A wide class of such operators is
constituted by the so-called weakly-nonlocal Hamiltonian operators \cite{MN01}:
\begin{equation}
\label{eq:103}
P^{ij} = P^{ij\sigma}\partial_\sigma  +
e^{\alpha}w^i_{\alpha} \partial_x^{-1} w^j_{\alpha},
\end{equation}
where $e^\alpha$ are constants and
$w^i_{\alpha} = w^i_{\alpha}(u^i,u^i_x,\ldots)$. The problem of defining the
Schouten bracket of nonlocal operators in a non-ambiguous way has caught the
interest of researchers, and it is still a lively topic. Indeed, Hamiltonian
operators have a precise geometric definition that allows to consider them as
geometric properties of PDEs, like symmetries or conserved quantities (see
\cite{KV11}). A rigorous approach to the computation of Schouten bracket for a
wide class of nonlocal Hamiltonian operators has been proposed in
\cite{DSK1,DSK2}, and it is based on the notion of nonlocal Poisson Vertex
Algebra. In the case of weakly nonlocal Hamiltonian operators a computational
solution to the problem has been recently proposed through the parallel
development of an algorithm in the three different languages of distributions,
operators, and Poisson Vertex Algebras \cite{CLV19}.

The Poisson bracket construction for local operators can be rephrased by means
of the well-known isomorphism between skew-symmetric linear differential
operators and superfunctions on jets of superbundles \cite{Getzler:DTHOpFCV}
(see also \cite{IVV,IgoninVerbovetskyVitolo:VMBGJS,KV11}). Basically, $n$ new
\emph{odd} dependent variables $p_1$,\dots, $p_n$ are introduced such that the
isomorphism reads on $P$ as
\begin{equation}
  \label{eq:6}
  P^{ij\sigma}\psi^1_i\partial_\sigma\psi^2_j \longrightarrow
  P^{ij\sigma} p_i p_{j\sigma},
\end{equation}
where $\psi^1$ and $\psi^2$ are the arguments of $P$ as a differential
operator.  Here the product $p_ip_{j\sigma}$ is the Grassmann product. Then,
the Schouten bracket between two operators can be written through a very
elegant formula:
\begin{equation}
  \label{eq:7}
  [P,Q] = \left[(-1)^{|P|}\fd{P}{u^i}\fd{Q}{p_i} +\fd{P}{p_i}\fd{Q}{u^i}\right].
\end{equation}
The square brackets on the right-hand side mean that the expression is
considered modulo total divergencies. This implies that in order to check that
the expression vanishes one should calculate its Euler--Lagrange operator and
see if the result is zero.

When writing \cite{CLV19} we considered the possibility to extend the results
of the paper to the formalism of superfunctions. Nonlocal terms in operators
can be represented by introducing new \emph{odd nonlocal variables}. Such a
technique was first introduced for Hamiltonian operators in
\cite{KerstenKrasilshchikVerbovetsky:HOpC}. In the case of weakly nonlocal
operators, this amounts at representing $P$ in \eqref{eq:103} as
\begin{equation}
  \label{eq:8}
  P = P^{ij\sigma}p_ip_{j\sigma}  +
e^{\alpha}w^i_{\alpha} p_i r_\alpha ,
\end{equation}
where $r_\alpha$ are new nonlocal \emph{odd} variables defined by
$\partial_x r_\alpha = w^j_{\alpha}p_j$.

Unfortunately, a naive rephrasing of the arguments of \cite{CLV19} in terms of
superfunctions does not work, as superfunctions do not allow to keep the role
of the arguments $\psi^1$, $\psi^2$, $\psi^3$ distinct in the three-vectors.

Weakly nonlocal Poisson operators associated with conformally flat metrics can
be interpreted as Jacobi structures. This beautiful observation allowed to find
a generalization of the formula \eqref{eq:7} to such operators in the framework
of supermanifolds (see \cite{LZ} for details).

In the present paper we consider general weakly nonlocal operators. Our
starting observation is that it is possible to extend the formula~\eqref{eq:7}
to weakly nonlocal superfunctions by a computational formula for the
variational derivatives of nonlocal odd variables.

We illustrate this extension through examples that cover a large class of
nonlocal Hamiltonian operators. More precisely, we prove the Hamiltonian
property for pseudodifferential operators for the Krichever-Novikov equation
and the modified KdV equation. Finally, we prove the well-known conditions that
are equivalent to the Hamiltonian property for an important subclass of
first-order weakly nonlocal operators introduced by Ferapontov \cite{Jenya}.

In the Conclusions we outline some features of this computational approach that
are promising in view of future investigations.

\section{Schouten bracket as Poisson bracket}
\label{sec:schouten-bracket-as}

We introduce new anticommuting (Grassmann) variables $p_i$,
$i=1$, \dots, $n$ and their $x$-derivatives $p_{i,\sigma}$. So, we work on the
(infinite order) jet of a superbundle with even coordinates $(x,u^i,u^i_\sigma)$
and odd coordinates $(p_i,p_{i,\sigma})$.

The formula \eqref{eq:7} for the variational Schouten bracket between two
operators, written as the superfunctions $P = P^{ij,\sigma}p_ip_{j,\sigma}$,
$Q = Q^{ij,\sigma}p_ip_{j,\sigma}$, makes use of the variational derivatives
\begin{equation}
  \label{eq:16}
  \fd{P}{u^i} = (-1)^{|\sigma|}\partial_\sigma\left(\pd{P}{u^i_\sigma}\right),
  \quad
  \fd{P}{p_i} = (-1)^{|\sigma|}\partial_\sigma\left(\pd{P}{p_{i\sigma}}\right),
\end{equation}
and similarly for $Q$.  Note that the derivatives with respect to odd
coordinates are odd derivatives, and total derivatives~\eqref{eq:8} are
extended to odd variables as
\begin{equation}
  \label{eq:12}
  \partial_\lambda = \pd{}{x^\lambda} + u^i_{\sigma,\lambda}\pd{}{u^i_\sigma}
  + p_{i\sigma,\lambda}\pd{}{p_{i\sigma}}.
\end{equation}
The square brackets on the right-hand side of~\eqref{eq:7} mean that we are in
an equivalence class: the expression in the bracket is considered up to total
divergencies of superfunctions. Thus, in order to check that $[P,Q]=0$ we need
to compute the variational derivative of the superfunction of degree $3$ inside
the square brackets at~\eqref{eq:7} and check that it is zero.

We would like to extend the Schouten bracket to weakly nonlocal operators. For
simplicity, we will only consider the case of one non-vanishing coefficient
$e^\alpha$; this means that we need to compute with superfunctions of the type
\begin{equation}
  \label{eq:54}
  P=P^{ij,\sigma}p_ip_{j,\sigma} + W^ip_ir,
\end{equation}
where $r$ is a superfunction of degree $1$ which is defined by the equation
\begin{equation}
  \label{eq:55}
  r_x = Z^jp_j,
\end{equation}
where $W$'s and $Z$'s are functions defined on the even part of the jet of the
superbundle. Note that in the case of just one nonlocal variable $r$ we have
$W=Z$ in order to guarantee skew-adjointness; but, as in more general cases we
might have several summands rearranged in such a way that $W\neq Z$, we prefer
to give a more general formula.

Since the Poisson bracket for superfunctions is graded-bilinear, in order to be
able to compute it for nonlocal superfunctions we just need a formula for the
variational derivative of the nonlocal part of the superfunction
\begin{equation}
  \label{eq:561}
  N=W^ip_ir.
\end{equation}
Let us introduce the notation $W=W^ip_i$ and
$Z=Z^ip_i$, and a new superfunction $s$ of degree $1$ defined
by the equation $s_x=W^ip_i$. Denote the Euler--Lagrange
operator by $\mathcal{E}$. It is well-known \cite{Verbovetsky:LFGAl} that if we
have a density $N=\int n\, dx$, and if we denote by $\ell_N$ its linearization
(or Fr\'echet derivative):
\begin{equation}
  \label{eq:9}
  \ell_N(\varphi)=\int\left(\pd{n}{u^i_\sigma}\partial_\sigma\varphi^i +
(-1)^{|n|+1}\pd{n}{p_{j,\sigma}}\partial_\sigma\varphi^j\right)\,dx
\end{equation}
then we have
\begin{equation}
  \label{eq:57}
  \mathcal{E}(N) = \ell^*_{N}(1) = \left(\fd{n}{u^i},\fd{n}{p_j}\right);
\end{equation}
note that signs on the odd part cancel after taking the adjoint. We also recall
the formula
$\ell_{\Delta_1\circ\Delta_2}(\varphi) = \ell_{\Delta_1}(\Delta_2(\varphi)) +
\Delta_1\circ\ell_{\Delta_2}(\varphi)$.

In order to compute the above expression, we also need the definition of the
adjoint operator of a superdifferential operator (\cite{Verbovetsky:LFGAl}; see
also \cite{IVV,IgoninVerbovetskyVitolo:VMBGJS}
\begin{equation}
  \langle q,\Delta(p)\rangle = (-1)^{|\Delta||q|}\langle \Delta^*(q),p\rangle
\end{equation}
where the angular brackets stand for duality in the variational cohomology
(i.e., the result is a density), and the absolute values have the value of the
parity of their arguments. We have
\begin{multline}
  \label{eq:58}
  \langle\ell^*_{N}(1),\varphi\rangle = \langle 1,\ell_N(\varphi)\rangle
  = \langle 1,\ell_{W,r}(\varphi) + W\ddx{-1}(\ell_Z(\varphi))\rangle
  \\
  = \langle \ell_{W,r}^*(1) + \ell^*_{Z,s}(1),\varphi\rangle.
\end{multline}
In the above formulae, expressions like $\ell_{W,r}(\varphi)$ are
linearizations of the term $W$ with fixed $r$.  Note that the parity of
$\ddx{-1}$ is $0$ and that ${\ddx{-1}}^*=-\ddx{-1}$.  In coordinates, we have
\begin{align}
  \label{eq:59}
  &\fd{N}{u^i} = (-1)^{|\sigma|}\partial_\sigma\left(\pd{W}{u^i_\sigma}r\right)
  + (-1)^{|\sigma|}\partial_\sigma\left(\pd{Z}{u^i_\sigma}s\right),
  \\
  &\fd{N}{p_i} = (-1)^{|\sigma|}\partial_\sigma\left(\pd{W}{p_{i,\sigma}}r\right)
  + (-1)^{|\sigma|}\partial_\sigma\left(\pd{Z}{p_{i,\sigma}}s\right),
\end{align}

The expression inside the square bracket at the right-hand side of~\eqref{eq:7}
is a superfunction of degree $3$; in order to check the vanishing of $[F,H]$ we
should be able to determine if the expression~\eqref{eq:7} is a total
divergence. So, we need to compute the Euler--Lagrange operator of a
$3$-superfunction with nonlocal terms. The only problems come from the nonlocal
terms. There might be two distinct types of nonlocal terms:
\begin{enumerate}
\item $T_1=Y_1^{i_1,\sigma_1;i_2,\sigma_2}p_{i_1,\sigma_1}p_{i_2,\sigma_2}r$;
  if we set
  $Y_1=Y_1^{i_1,\sigma_1;i_2,\sigma_2}p_{i_1,\sigma_1}p_{i_2,\sigma_2}$ we can
  proceed in a way which is similar to what we did for the nonlocal bivector
  $N$ (but keep into account the different gradings!), and get the formula
  \begin{equation}
    \label{eq:60}
    \ell_{T_1}^*(1) = \ell^*_{Y_1,r}(1) - \ell^*_{Z,y_1}(1).
  \end{equation}
  In coordinates:
  \begin{align}
    \label{eq:3}
    &\fd{T_1}{u^i} =
      (-1)^{|\sigma|}\partial_\sigma\left(\pd{Y_1}{u^i_\sigma}r\right)
  - (-1)^{|\sigma|}\partial_\sigma\left(\pd{Z}{u^i_\sigma}y_1\right),
  \\
    &\fd{T_1}{p_i} =
      (-1)^{|\sigma|}\partial_\sigma\left(\pd{Y_1}{p_{i,\sigma}}r\right)
  - (-1)^{|\sigma|}\partial_\sigma\left(\pd{Z}{p_{i,\sigma}}y_1\right),
  \end{align}
  where $y_{1,x}=Y_1$.
\item $T_2=Y_2^{i_1,\sigma_1}p_{i_1,\sigma_1}rs$; if we set
  $Y_2=Y_2^{i_1,\sigma_1}p_{i_1,\sigma_1}$ we have
  \begin{align}
    \label{eq:61}
    \langle\ell^*_{T_2}(1),\varphi\rangle =
    & \langle 1,\ell_{T_2}(\varphi) \rangle
    \\
    = & \langle 1,\ell_{Y_2,rs}(\varphi) + Y_2\ell_{r,s}(\varphi) +
      Y_2r\ell_s(\varphi)\rangle
    \\
    = & \langle 1,\ell_{Y_2,rs}(\varphi) + Y_2\ddx{-1}\ell_{Z,s}(\varphi) +
      Y_2r\ddx{-1}\ell_W(\varphi)\rangle
    \\
    = & \langle\ell_{Y_2,rs}^*(1) - \ell_{Z,sy_2}^*(1) - \ell_{W,y_3}^*(1)
        ,\varphi\rangle
  \end{align}
  where $y_2$ is defined by $(y_2)_x = Y_2$ and $y_3$ by $(y_3)_x = Y_2r$.
\end{enumerate}
We stress that the above approach is purely computational, as in the proof of
our formulae we make use of $\partial_x^{-1}$ which has no `good' geometrical
definition. However, we can show that in concrete computations our formula
reproduces known results; we hope to be able to provide a geometric
justification of the formula in the future.  Some new interesting developments
in this directon can be found in \cite{KV18}, where nonlocal operators are
treated in the framework of the geometry of jet spaces.

\section{Examples of computation}

In this Section we consider known examples of weakly nonlocal Hamiltonian
operators defining Poisson brackets. We provide a systematic computational approach to the
calculation of the Schouten bracket.

\subsection{Krichever--Novikov equation}

The Krichever--Novikov equation
\begin{equation}
  \label{eq:10}
  u_t = u_{xxx} - \frac{3}{2}\frac{u_{xx}^2}{u_x}
\end{equation}
has the Hamiltonian operator $P=u_x\d_x^{-1}u_x$ \cite{Sok84}. We rewrite it as
$P=N=Wr=u_xpr$, where $r_x=u_xp$. The Schouten bracket is
\begin{equation}
[N,N]=2\f{\delta N}{\delta u}\f{\delta N}{\delta p},\label{eq:11}
\end{equation}
where
\begin{equation}
\f{\delta N}{\delta u}=2\sum(-1)^k\d_x^k\left(\f{\delta W}{\delta
    u_{(k)}}r\right)=-2\d_x(pr)=-2p_xr-2p^2u_x=-2p_xr,\label{eq:13}
\end{equation}
and
\begin{equation}
\f{\delta N}{\delta p}=2\sum(-1)^k\d_x^k\left(\f{\delta W}{\delta
    p_{(k)}}r\right)=-2\d_x(pr)=2u_xr.\label{eq:14}
\end{equation}
Hence $[N,N]=-8u_xp_xr^2=0$ (no need to compute the variational derivative of
the $3$-superfunction in this simple case).

\subsection{Modified KdV}
\label{sec:modified-kdv}

The modified KdV equation is
\begin{equation}
  \label{eq:15}
  u_t = u^2u_x + u_{xxx};
\end{equation}
it has the weakly nonlocal Hamiltonian operator \cite{Wang02}
\begin{equation}
P=\d_x^3+\f{2}{3}u^2\d_x+\f{2}{3}uu_x-\f{2}{3}u_x\d_x^{-1}u_x\label{eq:17}.
\end{equation}

Let us set $P=L + N$, with $L=p_{xxx}p+\f{2}{3}u^2p_xp$ and $N=\f{2}{3}u_xpr$.
In the previous subsection we proved that $[N,N]=0$, hence the Schouten bracket
$[P,P]$ reduces to
\begin{equation}
  \label{eq:18}
  [L+N,L+N]=[L,L]+2[L,N]
\end{equation}
By a direct computation we have
\begin{equation}
[L,L]=\f{16}{3}upp_xp_{xxx}.\label{eq:19}
\end{equation}
Moreover
\begin{align*}
[L,&N]=\f{\delta L}{\delta u}\f{\delta N}{\delta p}+\f{\delta N}{\delta
        u}\f{\delta L}{\delta p}
  \\
  =&\left(\f{4}{3}up_xp\right)\left(\f{4}{3}u_xr\right)
     +\left(-\f{4}{3}p_xr\right)\left(-p_{xxx}-\f{2}{3}u^2p_x
     -\f{2}{3}\d_x(u^2p)-\d_x^3(p)\right)
  \\
  =&\f{16}{9}uu_xp_xpr-\f{4}{3}rp_xp_{xxx}+\f{16}{9}uu_xp_xrp
     +\f{4}{3}p_xrp_{xxx}=-\f{8}{3}rp_xp_{xxx}
     \end{align*}
Hence
\begin{equation}
[L+N,L+N]=\f{16}{3}(upp_x-rp_x)p_{xxx}.\label{eq:20}
\end{equation}
The above expression yields, after integrating it by parts:
\begin{multline}
  \f{16}{3}(upp_x-rp_x)p_{xxx}=-\f{16}{3}\d_x(upp_x-rp_x)p_{xx}
  \\
  =\f{16}{3}(-u_xpp_x+u_xpp_x)p_{xx}=0.\label{eq:1}
\end{multline}
More systematically, we can compute the Euler--Lagrange operator
of~\eqref{eq:20}. Let us set $T_L=upp_xp_{3x}$ and $T_N=-p_xp_{3x}r = $. We have:
\begin{align}
  \label{eq:2}
  \fd{T_L}{u} = &pp_xp_{3x}
  \\
  \fd{T_L}{p} = & -3u_{2x}pp_{2x} - 2u_xpp_{3x} - u_{3x}pp_x - 3u_xp_xp_{2x}
\end{align}
where the above computations have been done by the CDE package of Reduce
\cite{KVV17,reduce}. If $y_1$ is defined by $y_{1,x}=-p_xp_{3x}$ we observe
that, in this case $y_1= - p_xp_{2x}$. We have
\begin{align}
  \label{eq:4}
  \fd{T_N}{u} = &
  (-1)^{|\sigma|}\partial_\sigma \left(\pd{(-p_xp_{3x})}{u_\sigma}r\right)
  - (-1)^{|\sigma|}\partial_\sigma\left(\pd{(u_xp)}{u_\sigma}y_1\right)
  \\
  = & -\partial_x(pp_xp_{2x}) = - pp_xp_{3x},
\end{align}
and
\begin{align}
  \label{eq:42}
  \fd{T_N}{p} = &
  (-1)^{|\sigma|}\partial_\sigma \left(\pd{(-p_xp_{3x})}{p_\sigma}r\right)
  - (-1)^{|\sigma|}\partial_\sigma\left(\pd{(u_xp)}{p_\sigma}y_1\right)
  \\
  = & -\partial_x(-p_{3x}r) - \partial_{3x}(p_xr) - u_x(-p_xp_{2x})
  \\
  = & 3u_{2x}pp_{2x} + 2u_xpp_{3x} + u_{3x}pp_x + 3u_xp_xp_{2x}
\end{align}
which yields the result. We stress that, without the explicit integration of
$y_1$, the simplification would have not been possible.

\subsection{Example: first-order homogeneous weakly nonlocal
  operators}

The class of first-order homogeneous Poisson brackets was introduced in
\cite{DN}. This class is defined by first-order differential operators that are
homogeneous with respect to $x$-derivatives. The main feature of such operators
is that their `form' is preserved by coordinate transformations of the type
$\bar{u}^i = \bar{u}^i(u^j)$. This implies that the skew-symmetry and Jacobi
property of the Poisson brackets translate into geometric properties of the
coefficients of the differential operator.

The weakly nonlocal generalization of the above operators was introduced in a
very special case in \cite{FM90}, and later in a much wider sense in
\cite{Fer91}. The geometry of such operators is very rich and interesting; we
invite the reader to have a look at \cite{Jenya} and references therein.  We
will consider a weakly nonlocal first-order operator $P$ of the type:
\begin{equation}
  \label{eq:63}
  P^{ij} = g^{ij}\ddx{} + \Gamma^{ij}_k u^k_x
  +W^i_ku^k_x\ddx{-1}W^j_hu^h_x,
\end{equation}
although more general operators are possible and natural \cite{Jenya}.  Let us
introduce a nonlocal odd variable $r$ defined by $r_ x=W^j_hu^h_xp_j$, and
rewrite the operator $P$ in odd variables:
\begin{equation}
  \label{eq:65}
  P = g^{ij}p_{j,x}p_i +  \Gamma^{ij}_ku^k_xp_jp_i
  + W^i_ku^k_x\,r p_i
\end{equation}
Let us write $P=L+N$, where
\begin{equation}
  \label{eq:66}
  L=g^{ij}p_{j,x}p_i +  \Gamma^{ij}_ku^k_xp_jp_i,\qquad
  N=W^i_ku^k_x\,r p_i.
\end{equation}
We have to prove that the coefficients of $P$ satisfy the following
set of conditions (see \cite{Jenya})
\begin{subequations}\label{eq:611}
        \begin{gather}
        \label{eq:52}
        g^{ij} = g^{ji},
        \\
        \label{eq:56} g^{ij}_{,k} =
        \Gamma^{ij}_k+\Gamma^{ji}_k,
        \\
        \label{eq:572} g^{is}\Gamma^{jk}_s = g^{js}\Gamma^{ik}_s,
        \\
        \label{eq:582} g^{is}W^j_s = g^{js}W^i_s
        \\
        \label{eq:592} \nabla_i W^j_k = \nabla_k W^j_i,
        \\ \label{eq:602} R^{ij}_{kh} = W^i_kW^j_h - W^j_kW^i_h.
        \end{gather}
\end{subequations}
The skew-symmetry is equivalent to \eqref{eq:52} and \eqref{eq:56}, and is
assumed throughout the computation.

We use the formula
\begin{equation}
  \label{eq:67}
  [P,P] = [L+N,L+N] = [L,L] + 2[L,N] + [N,N].
\end{equation}
Then, from~\eqref{eq:7} it is clear that we need the formulae:
\begin{align}
  \label{eq:68}
  \fd{L}{u^l} = & 2\Gamma^{ji}_{l}p_{j,x}p_i +
                  (\Gamma^{ij}_{k,l} - \Gamma^{ij}_{l,k})u^k_x p_j p_i
  \\
  \fd{L}{p_l} = & - 2 g^{lj}p_{j,x}
                  + (\Gamma^{jl}_k - \Gamma^{lj}_k -g^{jl}_{,k})u^k_x p_j
                   =- 2 g^{lj}p_{j,x}-2\Gamma^{lj}_k\,u^k_x p_j
  \\
  \begin{split}
  \fd{N}{u^l} = &
    2(W^i_{l,k} - W^i_{k,l})
    u^k_x p_i\, r
    + 2W^i_l W^j_k u^k_x p_i p_j+ 2W^i_l p_{i,x} r
  \end{split}
  \\
  \fd{N}{p_l} =  & - 2W^{l}_k u^k_x \,r
\end{align}
Using~\eqref{eq:67} we compute three expressions; they are defined up to total
derivatives. We have:
\begin{eqnarray*}
  [L,L] &= & 2\fd{L}{u^i}\fd{L}{p_i}=\\
  &=&
      8\Gamma^{ji}_lg^{lm}p_ip_{j,x}p_{m,x}+\left(
      8\Gamma^{ji}_l\Gamma^{lm}_k
      +4g^{lj}(\Gamma^{im}_{k,l} - \Gamma^{im}_{l,k})\right)
      u^k_x p_{j,x}p_m p_i\\
  &&-4(\Gamma^{ij}_{h,l} - \Gamma^{ij}_{l,h})\Gamma^{lm}_k u^h_xu^k_x p_j p_ip_m
\end{eqnarray*}

\begin{eqnarray*}
  [L,N] &= & \fd{N}{u^i}\fd{L}{p_i} + \fd{L}{u^i}\fd{N}{p_i}=\\
  &=& \left(4g^{lj} (W^i_{l,k} - W^i_{k,l}) -4\Gamma^{li}_k
      W^j_l + 4 \Gamma^{ji}_{l}W^{l}_k\right)u^k_xp_i p_{j,x}\, r+\\
  && \left(4\Gamma^{lj}_h (W^i_{l,k} - W^i_{k,l})
    - 2W^{l}_k (\Gamma^{ji}_{h,l} - \Gamma^{ji}_{l,h})\right)
    u^h_xu^k_xp_i p_j \, r\\
  && - 4g^{lm}W^i_l W^j_k u^k_xp_i p_jp_{m,x}-4\Gamma^{lm}_h W^i_l W^j_k u^k_x u^h_x p_i p_j p_m + 4 g^{lj}W^i_l p_{i,x} p_{j,x}r
\end{eqnarray*}

Finally, taking into account that $r^2=0$ we obtain
\begin{equation*}
  [N,N] =  2\fd{N}{u^i}\fd{N}{p_i}= -8
  W^i_l W^j_k W^{l}_m u^k_xu^m_x p_i p_j\,r.
\end{equation*}

Collecting all the terms together we get
\begin{eqnarray*}
[P,P]&=&A^{ijh}p_ip_{j,x}p_{h,x}+B^{ijh}_ku^k_x p_ip_{j}p_{h,x}+C^{ijh}_{km}u^k_xu^m_x p_ip_{j}p_{h}+\\
&&D^{ij}_ku^k_xp_ip_{j,x}r+E^{ij}_{kh}u^k_xu^h_x p_ip_{j}r+F^{ij}_{hk}p_{i,x}p_{j,x}r
\end{eqnarray*}
that can be also written as
\begin{eqnarray*}
[P,P]&=&
\tilde{A}^{ijh}p_ip_{j,x}p_{h,x}+\tilde{B}^{ijh}_ku^k_x p_ip_{j}p_{h,x}+\tilde{C}^{ijh}_{km}u^k_xu^m_x p_ip_{j}p_{h}+\\
&&\tilde{D}^{ij}_ku^k_xp_ip_{j,x}r+\tilde{E}^{ij}_{kh}u^k_xu^h_x p_ip_{j}r+\tilde{F}^{ij}p_{i,x}p_{j,x}r
\end{eqnarray*}
with
\begin{eqnarray*}
\tilde{A}^{ijh}&=&\f{1}{2}(A^{ijh}-A^{ihj}),\\
\tilde{B}^{ijh}_k&=&\f{1}{2}(B^{ijh}_k-B^{jih}_k)\\
\tilde{C}^{ijm}_{hk}&=&\f{1}{12}(C^{ijm}_{kh}-C^{imj}_{kh}-C^{jim}_{kh}+C^{jmi}_{kh}+C^{mij}_{kh}-C^{mji}_{kh})+\\
&&\f{1}{12}(C^{ijm}_{hk}-C^{imj}_{hk}-C^{jim}_{hk}+C^{jmi}_{hk}+C^{mij}_{hk}-C^{mji}_{hk})\\
\tilde{D}^{ij}_k&=&D^{ij}_k,\\
\tilde{E}^{ij}_{kh}&=&\f{1}{4}(E^{ij}_{kh}-E^{ji}_{kh})+\f{1}{4}(E^{ij}_{hk}-E^{ji}_{hk}),\\
\tilde{F}^{ij}&=&\f{1}{2}(F^{ij}-F^{ji})
\end{eqnarray*}

We obtain
\begin{eqnarray*}
  \tilde{A}^{ijh}&=&4
                     (g^{lh}\Gamma^{ji}_l-g^{lj}\Gamma^{hi}_l)
  \\
  \tilde{B}^{ijh}_k&=&4R^{ijh}_k+4g^{li}(W^h_kW^j_l-W^j_kW^h_l)
  \\
  \tilde{C}^{ijm}_{hk}&=
  &-\f{2}{3}\Gamma^{lm}_k( R^{ij}_{lh}-W^i_lW^j_h+W^j_lW^i_h)
    -\f{2}{3}\Gamma^{lm}_h( R^{ij}_{lk}-W^i_lW^j_k+W^j_lW^i_k)
  \\
&&+\f{2}{3}\Gamma^{li}_k(
   R^{mj}_{lh}-W^m_lW^j_h+W^j_lW^m_h)+\f{2}{3}\Gamma^{li}_h(
   R^{mj}_{lk}-W^m_lW^j_k+W^j_lW^m_k)
  \\
&&-\f{2}{3}\Gamma^{lj}_k(
   R^{mi}_{lh}-W^m_lW^i_h+W^i_lW^m_h)-\f{2}{3}\Gamma^{lj}_h(
   R^{mi}_{lk}-W^m_lW^i_k+W^i_lW^m_k)
  \\
  \tilde{D}^{ij}_k&=&2g^{lj}(\nabla_kW^i_l-\nabla_lW^i_k)
                      -2\Gamma^i_{mk}(g^{jl}W^m_l-g^{ml}W^j_l)
  \\
  \tilde{E}^{ij}_{hk}&=&2\Gamma^{lj}_h(\nabla_k W^i_l-\nabla_l
                                        W^i_k)-2\Gamma^{li}_h(\nabla_k
                                        W^j_l-\nabla_l W^j_k)+
  \\
&&2\Gamma^{lj}_k(\nabla_h W^i_l-\nabla_l W^i_h)-2\Gamma^{li}_k(\nabla_h
   W^j_l-\nabla_l W^j_h)+
  \\
&&W^l_k(
   R^{ij}_{lh}-W^i_lW^j_h+W^j_lW^i_h)+W^l_h(
   R^{ij}_{lk}-W^i_lW^j_k+W^j_lW^i_k)
  \\
&&-W^l_k(
   R^{ji}_{lh}-W^j_lW^i_h+W^i_lW^j_h)-W^l_h(
   R^{ji}_{lk}-W^j_lW^i_k+W^i_lW^j_k)
  \\
\tilde{F}^{ij}&=&2(g^{lj}W^{i}_l-g^{li}W^{j}_l)
\end{eqnarray*}
Let us set $T=[P,P]$. The system
\begin{equation}
  \label{eq:21}
  \fd{T}{u^l} = 0,\qquad \fd{T}{p_l} = 0,
\end{equation}
yields the following conditions:
\begin{itemize}
\item $\tilde{B}^{ijh}_l = 0$, which is the coefficient of $p_ip_jp_{h,2x}$ in
  $\fd{T}{u^l}$;
\item $2\tilde{A}^{ilj} = 0$, which is the coefficient of $p_ip_{j,2x}$ in
  $\fd{T}{p_l}$;
\item $- \tilde{D}^{il}_k =0$, which is the coefficient of $u^k_{2x}p_ir$ in
  $\fd{T}{p_l}$;
\item $2\tilde{F}^{il} = 0$, which is the coefficient of $p_{i,2x}r$ in
  $\fd{T}{p_l}$.
\end{itemize}
The above conditions are equivalent to the conditions~\eqref{eq:572},
\eqref{eq:582}, \eqref{eq:592}, \eqref{eq:602}, and
imply the vanishing of the coefficients $\tilde{C}$ and $\tilde{E}$.

We remark that the last step of the computation of the Schouten bracket is
very straightforward: it is easy to derive the vanishing conditions from few
selected coefficients in the variational derivative.

\section{Conclusions}

Weakly non local nonlocal hamiltonian operators arise naturally in the theory
and applications of integrable systems \cite{MN01}.

In \cite{CLV19} we developed an algorithm to compute Schouten brackets of such
operators using three different formalisms: distributions, pseudodifferential
operators, Poisson vertex algebras. In this paper we propose an alternative
approach based on the identification of weakly non local hamiltonian operators
with superfunctions on supermanifolds. This approach requires to define
variational derivative for nonlocal variables. This allows to extend the known
formula for the Schouten bracket of local operators in a straightforward way.

Finding necessary and sufficient conditions for the vanishing of the bracket is
not immediate, the main difficulty being that the nonlocal odd variables that
arise in the computations should be checked in order to see if they can be
integrated (see Section 2.2).  However this problem can be easily fixed and the
implementation of the main result on a computer algebra program seems possible.
For instance, the Reduce package \cite{KVV17,Vit19} already allows to use local
and nonlocal variables and contains an implementation of the Schouten bracket
for local operators in terms of odd variables.

A set of software packages for the symbolic calculation of the Schouten bracket
adapted to all the above formalisms will be the subject of our future work.

\end{document}